\documentclass[10pt]{article}

\usepackage[utf8]{inputenc}
\usepackage[T1]{fontenc}
\usepackage{arxiv}
\usepackage{url}
\usepackage{booktabs}
\usepackage{tabularx}
\usepackage{multirow}
\usepackage{chemformula}
\usepackage{float}
\usepackage{caption}
\usepackage{amsmath, amssymb, amsfonts}
\usepackage{graphicx}
\usepackage{xcolor}
\usepackage[numbers,sort&compress]{natbib}
\usepackage{algorithm}
\usepackage{algpseudocode}
\usepackage{xspace}
\usepackage{microtype}
\usepackage{hyperref}
\newcolumntype{P}[1]{>{\centering\arraybackslash}p{#1}}
\begin{document}

\title{A Systematic Benchmark of Physics-Informed Neural Network
  Architectures for the Stiff Poisson--Nernst--Planck System:
  Adaptive Loss Weighting and Multi-Scale Resolution}

\author{%
  \begin{minipage}[t]{0.48\linewidth}
    \centering
    \textbf{David Pankaczy}$^{1}$\\[3pt]
    {\small $^{1}$Department of Physics and Astronomy\\
            University of Waterloo\\
            200 University Ave.\ West\\
            Waterloo, ON N2L\,3G1, Canada}\\[2pt]
    {\small \texttt{dpankacz@uwaterloo.ca}}
  \end{minipage}\hfill
  \begin{minipage}[t]{0.48\linewidth}
    \centering
    \textbf{Conrard Giresse Tetsassi Feugmo}$^{1,2,*}$\\[3pt]
    {\small $^{1}$Department of Physics and Astronomy\\
            $^{2}$Department of Chemistry\\
            University of Waterloo\\
            200 University Ave.\ West\\
            Waterloo, ON N2L\,3G1, Canada}\\[2pt]
    {\small $^{*}$\texttt{cgtetsas@uwaterloo.ca}}
  \end{minipage}%
}
\date{}

\maketitle

\begin{abstract}

The Poisson--Nernst--Planck (PNP) system constitutes a canonical stiff coupled
PDE problem: the charge-density prefactor $F/\varepsilon_0 \approx 10^{16}$\,C\,V$^{-1}$\,m$^{-3}$
produces extreme coefficient ratios, the electric double layer imposes sharp
boundary layers with a singular-perturbation character, and the Poisson and
Nernst--Planck equations are nonlinearly coupled.
Physics-informed neural networks (PINNs) are appealing here because they require no
mesh, differentiate through the physics automatically, and handle forward and inverse
problems in one framework. Spectral bias and multi-task loss imbalance, however, have
limited their accuracy on stiff PNP systems.
We present the first systematic, data-free benchmark of eleven PINN configurations,
organised into four strategy groups (adaptive loss weighting, spectral bias mitigation,
spatio-temporal decomposition, and physics enrichment), on a physically parametrised
one-dimensional PNP model for a lithium symmetric cell, implemented entirely within
NVIDIA PhysicsNeMo Sym and validated against a finite volume method (FVM) reference.
Root-mean-square errors (RMSE) span $10^{-2}$--$10^{-4}$ across architectures:
Neural Tangent Kernel (NTK) adaptive loss weighting achieves
RMSE\,$= 6.6\times10^{-4}$ (anion), $6.2\times10^{-4}$ (cation), and
$1.1\times10^{-3}$ (electric potential).
The balanced residual decay rate (BRDR) scheme matches NTK within 10\% of RMSE
for the concentration fields (24\% for the electric potential)
while reducing mean wall-clock time by $3.2 \pm 0.4$\,h per run, making it the
preferable strategy under compute constraints.
Loss landscape geometry corroborates the RMSE ranking: NTK yields the sharpest,
most symmetric basin, while poorly conditioned architectures such as Separable PINNs (SPINNs) exhibit
flat, irregular landscapes.
We release an open-source PhysicsNeMo Sym implementation for reuse on stiff coupled
PDE problems in computational mechanics.
\end{abstract}


\section{Introduction}
\label{sec:intro}

The Poisson--Nernst--Planck (PNP) system is a prototypical example of a stiff,
nonlinearly coupled PDE problem that challenges classical and modern numerical
solvers alike.
The dimensionless form of the system, derived in Section~\ref{sec:model}, admits
a singular-perturbation structure controlled by the parameter
$\varepsilon \propto \lambda_D/L$, where $\lambda_D$ is the Debye screening length
and $L$ is the domain length.
For typical electrochemical parameters, $\varepsilon \ll 1$ ($\varepsilon \approx
10^{-4}$--$10^{-5}$ for concentrated electrolytes), which forces sharp boundary
layers at the electrode interfaces that classical finite element codes address
through Slotboom variable transformations, Gummel-iteration decoupling, and
locally refined meshes~\cite{XIE2020109915}.
At the same time, the extreme charge-density prefactor $F/\varepsilon_0 \approx
1.09\times10^{16}$\,C\,V$^{-1}$\,m$^{-3}$ creates coefficient ratios that
destabilize standard iterative solvers~\cite{Bazant2004}.
The physical application motivating this work is ion transport in a one-dimensional
lithium symmetric cell with \ch{LiPF6} electrolyte, which provides a parametrically
well-documented and experimentally validated instance of the PNP system
\cite{Subramariam2019,Wood2016-yf,Newman2004}, and the same equations govern
ion channels~\cite{Eisenberg1996}, nanofluidic devices~\cite{Daiguji2004}, and
solid-state ionic transport~\cite{Bazant2004}.

Neural networks for solving PDEs were first proposed by \citet{Lagaris1998ANN};
the modern physics-informed neural network (PINN) framework of \citet{RAISSI2019686}
placed this on a rigorous automatic differentiation footing and has since become a
leading paradigm in scientific machine learning
\cite{Karniadakis2021review,Cuomo2022,EYu2018deep}.
PINNs are attractive for stiff coupled PDEs because they require no mesh, differentiate
through physical laws automatically, and accommodate operator-learning extensions
\cite{Wang2021DeepONets}.
However, two fundamental difficulties limit their accuracy on PNP-class problems.
The first is \emph{spectral bias}: PINNs preferentially learn low-frequency
components of the solution, making the stiff, higher-frequency Poisson equation
difficult to resolve~\cite{rahaman2019spectralbias,xu2019fprinciple,Krishnapriyan2021failure}.
The second is \emph{multi-task loss imbalance}: because the PNP system couples
equations with vastly different characteristic scales, the individual loss
components converge at different rates, and naive uniform weighting causes the
optimizer to over-satisfy the smoother Nernst--Planck equations at the expense
of the stiffer Poisson equation
\cite{WANG2022110768,Wang2021understanding,Wang2024causality}.
Rigorous error bounds for PINNs on elliptic and parabolic PDEs have been established
by \citet{Shin2020convergence} and generalisation error estimates by
\citet{Mishra2022estimates,Mishra2023forward,DeRyck2022PINNerror}; however, none of
these analyses address the regime of extreme coefficient ratios ($\varepsilon^2 \sim
10^{-10}$) characteristic of the PNP system studied here.

Prior PINN work on PNP has not adequately resolved either difficulty at
practically useful accuracy.
\citet{HUANG2025231} proposed data-free enriched PINNs for dynamic PNP systems
and demonstrated accuracy improvements over vanilla PINNs, but did not conduct a
systematic multi-architecture benchmark under battery-relevant parametrisation.
No systematic, data-free, multi-architecture benchmark exists for the PNP system
under battery-relevant parametrisation.

This paper closes that gap.
We benchmark eleven PINN configurations, organised into four strategy groups, against
a validated FVM reference on the one-dimensional PNP system of \citet{Subramariam2019},
implemented entirely within NVIDIA's open-source PhysicsNeMo Sym framework~\cite{physicsnemo2024}.
The four groups are: \emph{adaptive loss weighting} (NTK, BRDR, AdaHessian), which corrects
gradient pathologies without changing the network architecture; \emph{spectral bias mitigation}
(Fourier features, PIKAN), which modifies the network's input representation or basis functions
to resolve high-frequency boundary layers; \emph{spatio-temporal decomposition} (FBPINN,
Decoupled, SPINN, Sym./antisym.\ transform), which reduces problem complexity by splitting the
domain, the equations, or the solution variables; and \emph{physics enrichment} (EPINN),
which embeds stiff analytical features directly into the network basis.
The specific contributions are as follows.
We provide the first systematic, data-free benchmark of eleven PINN configurations spanning
four methodological groups on a physically parametrised 1D PNP system, with errors averaged
over ten independent training runs per configuration.
We show that NTK adaptive loss weighting achieves RMSE as low as $6.6\times10^{-4}$
(dimensionless), while BRDR weighting matches this within 9\% at reduced
computational cost.
A loss landscape analysis provides geometric corroboration of the RMSE ranking;
wall-time distributions characterise all eleven configurations on an NVIDIA H100 GPU.
We also release an open-source, modular PhysicsNeMo Sym implementation for PNP
problems and, more broadly, for stiff coupled PDE systems in computational electrochemistry.

\section{Developments in Physics-Informed Neural Networks}
\label{sec:pinn_review}

Since \citet{RAISSI2019686} introduced the modern PINN framework, spectral bias
\cite{rahaman2019spectralbias} and multi-task loss imbalance have been the main
obstacles, and both are especially acute for the PNP system studied here.
Spectral bias arises because standard multilayer perceptrons (MLPs) preferentially
fit low-frequency, global trends, leaving high-frequency localised features such
as the electric double layer (EDL) unresolved.
Loss imbalance arises because stiff multiphysics systems impose loss components
with disparate magnitudes and convergence rates, so that a uniform weighting
neglects the stiffest equations.

\citet{Tancik2020FourierFeatures} show that Fourier feature mappings applied to
the input layer enable learning of high-frequency components for low-dimensional
problems; \citet{Wang2021eigen} extend this by analysing the eigenvalue structure
of the resulting NTK and proposing modified Fourier networks.
\citet{Wang2021understanding} identify gradient flow pathologies as the root
cause of PINN training failure on multiphysics problems, and
\citet{WANG2022110768} develop NTK-based adaptive loss weighting, grounded in
the infinite-width NTK theory of \citet{Jacot2018NTK}, which equates convergence
rates across loss components and substantially improves accuracy on multi-scale and
nonlinear systems.
Domain-decomposition approaches, originally proposed as conservative PINNs by
\citet{Jagtap2020conservative} and extended to finite basis PINNs (FBPINNs)
by \citet{Moseley2023FBPINNs} with multilevel decomposition by
\citet{DOLEAN2024117116}, address both spectral bias and the
global-communication bottleneck between subdomains.
Separable PINNs~\cite{Cho2023SPINNs} reduce memory requirements by canonical polyadic (CP) tensor
decomposition but sacrifice accuracy on problems where coupling between spatial
dimensions is strong.
Kolmogorov--Arnold networks (KANs)~\cite{liu2025kan}, implemented for PDE solving
as PIKANs by \citet{WANG2025PIKAN}, offer theoretical advantages over MLPs but
exhibit increased inference cost due to basis function computation.
\citet{McClenny2023selfadaptive} introduce per-collocation-point trainable weights
(self-adaptive PINNs) as a further adaptive strategy.
\citet{CHEN2025BRDR} propose the balanced residual decay rate (BRDR) method,
demonstrating accuracy comparable to NTK weighting at lower computational cost.
Second-order optimisation via AdaHessian~\cite{Yao2021AdaHessian} provides
curvature information through Hutchinson's estimator, offering an alternative
convergence path.
Our benchmark evaluates all of these strategies on the same PNP problem, providing
the first head-to-head comparison on a stiff, nonlinearly coupled system with
physical parametrisation.

\section{Poisson--Nernst--Planck Model}
\label{sec:model}

\subsection{Dimensional PNP system}
\label{sec:dim_pnp}

\begin{figure}[H]
    \centering
    \includegraphics[width=0.95\textwidth]{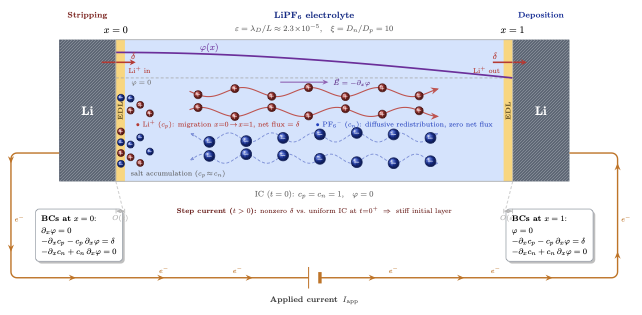}

    	\caption{\textbf{Schematic of the one-dimensional lithium symmetric cell.}
    		The computational domain ($x\in[0,1]$) spans the \ch{LiPF6} electrolyte
    		between two identical lithium metal electrodes.
    		Under the applied positive current density $I_\mathrm{app}$, the cation
    		(\ch{Li+}, $c_p$) migrates toward the cathode ($x=1$) driven by the electric
    		field.
    		The anion (\ch{PF6-}, $c_n$) carries zero net flux at both electrode
    		interfaces (no-flux boundary condition); it accumulates near $x=0$ as
    		electromigration and diffusion balance inside the electric double layer.
    		The electric potential $\varphi$ is the third solved field.
    		Thin bands at both interfaces represent the electric double layer of
    		dimensionless thickness $O(\varepsilon)\approx 2.3\times10^{-5}$.\label{fig:cell_schematic}}

\end{figure}

Our base-case system is the one-dimensional PNP transport model of
\citet{Subramariam2019}, which describes the mean-field dynamics of a binary
electrolyte (\ch{LiPF6}) in a lithium symmetric cell.
The model solves for the cation concentration $C_p$ (\ch{Li+}), anion
concentration $C_n$ (\ch{PF6-}), and electric potential $\Phi$.
The dimensional governing equations, boundary conditions (BCs), and initial
condition (IC) are collected in Table~\ref{tab:dim_pnp}; physical parameters are
given in Table~\ref{tab:params}.
The cell geometry is illustrated in Figure~\ref{fig:cell_schematic}.

\begin{table}[H]
    \hrule\vspace{0.15cm}
    \caption{\textbf{Dimensional PNP system.}
    Interior PDE, boundary conditions (BCs), and initial condition (IC) for the
    cation concentration $C_p$, anion concentration $C_n$, and electric potential
    $\Phi$. Both electrode interfaces impose identical flux BCs; the right BC for
    each field is identical in form to the left BC.\label{tab:dim_pnp}}
    \renewcommand{\arraystretch}{1.6}%
    \setlength{\tabcolsep}{4pt}%
    \begin{tabular}{p{6.6cm}p{5.0cm}p{1.9cm}p{1.5cm}}
        \textbf{Interior PDE} & \textbf{Left BC ($X=0$)}
          & \textbf{Right BC} & \textbf{IC}\\
    \midrule
        $\displaystyle\frac{\partial^2 \Phi}{\partial X^2}
          = -\frac{F}{\varepsilon_0\varepsilon_s}(z_p C_p + z_n C_n)$
        & $\displaystyle\frac{\partial\Phi}{\partial X}=0$
        & $\Phi=0$ & $\Phi=0$\\
        $\displaystyle\frac{\partial C_p}{\partial \mathcal{T}}
          = D_p\frac{\partial^2 C_p}{\partial X^2}
          + \frac{z_p D_p F}{RT}\frac{\partial}{\partial X}\!\left(C_p\frac{\partial\Phi}{\partial X}\right)$
        & $\displaystyle-D_p\frac{\partial C_p}{\partial X}
          -\frac{z_p D_p F}{RT}C_p\frac{\partial\Phi}{\partial X}
          =\frac{I_\mathrm{app}}{z_p F}$
        & same as left
        & $C_p=C_0$\\
        $\displaystyle\frac{\partial C_n}{\partial \mathcal{T}}
          = D_n\frac{\partial^2 C_n}{\partial X^2}
          + \frac{z_n D_n F}{RT}\frac{\partial}{\partial X}\!\left(C_n\frac{\partial\Phi}{\partial X}\right)$
        & $\displaystyle-D_n\frac{\partial C_n}{\partial X}
          -\frac{z_n D_n F}{RT}C_n\frac{\partial\Phi}{\partial X}=0$
        & same as left
        & $C_n=C_0$\\
    \end{tabular}
    \renewcommand{\arraystretch}{1}%
    \setlength{\tabcolsep}{6pt}%
    \vspace{0.1cm}\hrule
\end{table}

\begin{table}[H]
    \hrule\vspace{0.15cm}
    \caption{\textbf{Physical parameters for \ch{LiPF6} in a lithium symmetric
    cell}~\cite{Subramariam2019}.
    The diffusivity ratio $D_n/D_p = 10$ and the small Debye-to-domain-length
    ratio $\varepsilon \approx 2.3\times10^{-5}$ are the primary sources of
    stiffness.\label{tab:params}}
    \begin{tabularx}{\textwidth}{XXXP{5.5cm}}
        \textbf{Symbol} & \textbf{Value} & \textbf{Units} & \textbf{Description}\\
    \midrule
        $C_0$           & $500$               & mol\,m$^{-3}$   & Initial electrolyte concentration\\
        $D_p$           & $4\times10^{-10}$   & m$^2$\,s$^{-1}$ & Diffusivity of \ch{Li+}\\
        $D_n$           & $4\times10^{-9}$    & m$^2$\,s$^{-1}$ & Diffusivity of \ch{PF6-}\\
        $z_p$           & $1$                 & ---             & Cation charge number\\
        $z_n$           & $-1$                & ---             & Anion charge number\\
        $I_\mathrm{app}$& $7.72$              & A\,m$^{-2}$     & Applied current density (chosen to give $\delta=0.3$)\\
        $\varepsilon_s$ & $16.8$              & ---             & Relative permittivity of solvent\\
        $T$             & $298.15$            & K               & Temperature\\
        $L$             & $7.5\times10^{-4}$  & m               & Inter-electrode distance\\
        $\mathcal{T}_f$ & $3\,600$            & s               & Final simulation time\\
        $\lambda_D$     & $\approx 1.7\times10^{-8}$ & m       & Debye length ($\varepsilon \approx 2.3\times10^{-5}$)\\
    \end{tabularx}
    \vspace{0.1cm}\hrule
\end{table}

\subsection{Nondimensionalisation and conditioning}
\label{sec:nondim}

Nondimensionalisation is essential for PINN training: dimensional variables span
orders of magnitude that produce ill-conditioned loss functions and impede
convergence~\cite{Haghighat2021,Markidis2021}.
We adopt the scheme of \citet{Subramariam2019}:
\begin{gather}
    c_p = C_p / C_0, \quad c_n = C_n / C_0, \quad \varphi = \Phi F/(RT)
    \label{eq:ndim1}\\
    x = X/L, \quad t = \mathcal{T}\, D_p / L^2
    \label{eq:ndim2}\\
    \varepsilon = \sqrt{\frac{RT\varepsilon_0\varepsilon_{s}}{z_p^2 F^2 C_0 L^2}},
    \quad \delta = \frac{I_\mathrm{app}}{z_p F C_0 D_p / L}, \quad \xi = \frac{D_n}{D_p}
    \label{eq:ndim3}
\end{gather}
where $\varepsilon$ is the ratio of the Debye screening length $\lambda_D$ to the
domain length $L$ (the singular-perturbation parameter), $\delta$ is the
dimensionless applied current, and $\xi = D_n/D_p = 10$ is the diffusivity ratio.
For the parameters of Table~\ref{tab:params} these evaluate to
$\varepsilon \approx 2.30\times10^{-5}$ and $\delta = 0.3$;
$I_\mathrm{app}$ is set to $7.72$\,A\,m$^{-2}$ (slightly below the
$10$\,A\,m$^{-2}$ of~\cite{Subramariam2019}) so that $\delta = 0.3$ serves
as the canonical dimensionless current for all PINN and FVM computations
in this benchmark.
The dimensionless system is given in Table~\ref{tab:nondim_pnp}.

The parameter $\varepsilon \approx 2.30\times10^{-5}$ is small, which means the
Poisson equation $-\varepsilon^2 \varphi_{xx} = z_p c_p + z_n c_n$ has a
coefficient $\varepsilon^2 \approx 5.3\times10^{-10}$: the electric potential
must satisfy this equation with a coefficient ten orders of magnitude smaller than
unity, creating the loss imbalance that is the central computational difficulty
for all PINN configurations tested.

\begin{table}[H]
    \hrule\vspace{0.15cm}
    \caption{\textbf{Dimensionless PNP system.}
    Derived from the nondimensionalisation
    ~\eqref{eq:ndim1}--\eqref{eq:ndim3}.
    The small parameter $\varepsilon \approx 2.3\times10^{-5}$ controls the
    sharpness of the electric double layer at $x=0$ and $x=1$.\label{tab:nondim_pnp}}
    \begin{tabularx}{\textwidth}{p{5.8cm}XXX}
        \textbf{Interior} & \textbf{Left BC} & \textbf{Right BC} & \textbf{IC}\\
    \midrule
        $\displaystyle-\varepsilon^2\frac{\partial^2\varphi}{\partial x^2}
          = z_p c_p + z_n c_n$
        & $\displaystyle\frac{\partial\varphi}{\partial x}=0$
        & $\varphi=0$ & $\varphi=0$\\[10pt]
        $\displaystyle\frac{\partial c_p}{\partial t}
          = \frac{\partial^2 c_p}{\partial x^2}
          + z_p\frac{\partial}{\partial x}\!\left(c_p\frac{\partial\varphi}{\partial x}\right)$
        & $\displaystyle-\frac{\partial c_p}{\partial x}
          - z_p c_p\frac{\partial\varphi}{\partial x} = \delta$
        & $\displaystyle-\frac{\partial c_p}{\partial x}
          - z_p c_p\frac{\partial\varphi}{\partial x} = \delta$
        & $c_p=1$\\[10pt]
        $\displaystyle\frac{\partial c_n}{\partial t}
          = \xi\!\left[\frac{\partial^2 c_n}{\partial x^2}
          + z_n\frac{\partial}{\partial x}\!\left(c_n\frac{\partial\varphi}{\partial x}
            \right)\right]$
        & $\displaystyle-\frac{\partial c_n}{\partial x}
          - z_n c_n\frac{\partial\varphi}{\partial x}=0$
        & $\displaystyle-\frac{\partial c_n}{\partial x}
          - z_n c_n\frac{\partial\varphi}{\partial x}=0$
        & $c_n=1$\\
    \end{tabularx}
    \vspace{0.1cm}\hrule
\end{table}

\subsection{Finite volume reference solver}
\label{sec:fvm}

The reference solution is obtained by a \emph{method of lines} (MOL) solver
implemented in Python (\texttt{pnp\_fvm.py}).
The spatial domain $x\in[0,1]$ is discretised on a uniform grid of
$N_\mathrm{FVM} = 500$ cell-centred nodes with spacing $h = 1/N_\mathrm{FVM}
= 2\times10^{-3}$.
At each time step the Poisson equation in Table~\ref{tab:nondim_pnp} is solved algebraically
as a $(N_\mathrm{FVM}+1)\times(N_\mathrm{FVM}+1)$ tridiagonal linear system
(\texttt{numpy.linalg.solve}), with the left Neumann condition
$\partial_x\varphi|_{x=0}=0$ discretised by a first-order one-sided difference
and the right Dirichlet condition $\varphi(1,t)=0$ imposed directly.
Ion fluxes are approximated at cell faces using centered differences for
concentration gradients and arithmetic-average face concentrations, consistent
with the standard cell-centred finite-volume flux:
\begin{equation}
  N_p^{(j+\frac{1}{2})} = -\frac{c_p^{(j+1)}-c_p^{(j)}}{h}
    - \bar{c}_p^{(j+\frac{1}{2})}
      \frac{\varphi^{(j+1)}-\varphi^{(j)}}{h},
  \label{eq:fvm_flux}
\end{equation}
and analogously for $N_n$.
The resulting stiff ODE system for the $2(N_\mathrm{FVM}+1)$ concentration
variables is integrated in time by \texttt{scipy.integrate.solve\_ivp} with
the \texttt{Radau} implicit Runge--Kutta solver (order~5,
tolerances $\mathtt{rtol}=10^{-6}$, $\mathtt{atol}=10^{-8}$), which is
well-suited to the stiff eigenvalue spectrum arising from the $\varepsilon^{-2}$
coefficient in the Poisson equation.
The boundary flux is set to $\delta = 0.3$ ($I_\mathrm{app} \approx 7.72$\,A\,m$^{-2}$),
matching the value used in all PINN configurations.
The solution is evaluated at $3\,600$ uniformly spaced times on
$[0,\,\tau_f\approx 2.56]$, yielding a space-time dataset of
$3\,600\times 501$ points for each of the three fields ($c_p$, $c_n$, $\varphi$)
written to \texttt{pnp.csv}.
The spatial discretisation error, estimated by Richardson extrapolation between
$N=250$ and $N=500$ grids, is $O(h^2)\approx 4\times10^{-6}$,
at least two orders of magnitude below the best PINN RMSE ($6.6\times10^{-4}$),
so the MOL solution is a reliable benchmark reference.
A full derivation of the discrete equations, boundary conditions, and solver
parameters is given in Supplementary~Material~Section~1.
Existence and uniqueness of the continuous PNP solution for the parameter regime
studied here follows from \citet{Jerome1996analysis}.

\section{PINN Benchmark Design and Training Configurations}
\label{sec:method}

\subsection{Physics-informed neural networks}
\label{sec:pinn_formulation}

Universal approximation theorems~\cite{Hornik1989,Cybenko1989} guarantee that a
sufficiently expressive neural network can approximate any continuous function to
arbitrary precision.
PINNs exploit this by embedding physical laws directly into the loss function via
automatic differentiation (AD), which removes the need for a mesh.
Formally, the PINN problem~\eqref{eq:prob}--\eqref{eq:update} is:
\begin{gather}
    \theta^\ast = \arg \min_{\theta} \mathcal{L}(\theta) \label{eq:prob} \\
    \theta_{n+1} \gets \theta_n - \alpha_n \nabla_{\theta} \mathcal{L}(\theta_n)
    \label{eq:update}
\end{gather}
where $\theta$ denotes network parameters, $\alpha_n$ is the learning rate at
iteration $n$, and Adam~\cite{Kingma2014} provides adaptive per-parameter learning
rates.

The composite loss for the PNP system is:
\begin{equation}
    \mathcal{L}(\theta) = \lambda_{\mathrm{PDE}}\,\mathcal{L}_{\mathrm{PDE}}
                + \lambda_{\mathrm{BC}}\,\mathcal{L}_{\mathrm{BC}}
                + \lambda_{\mathrm{IC}}\,\mathcal{L}_{\mathrm{IC}}
    \label{eq:loss}
\end{equation}
where $\mathcal{L}_{\mathrm{PDE}}$, $\mathcal{L}_{\mathrm{BC}}$, and
$\mathcal{L}_{\mathrm{IC}}$ are the mean-squared PDE residual, boundary condition
residual, and initial condition residual, respectively.
The weights $\lambda_i$ are either fixed at unity (vanilla) or adaptively
determined at each training step by one of the two adaptive schemes described
in Section~\ref{par:weighting}.

\subsection{PhysicsNeMo Sym framework}
\label{sec:physicsnemo}

All implementations use NVIDIA's open-source PhysicsNeMo Sym framework
\cite{physicsnemo2024}, formerly NVIDIA Modulus Sym, built on PyTorch.
The framework has four features that are central to the benchmark design.

\textbf{Computational graph and automatic differentiation.}
Computations are organised as a directed acyclic graph of \texttt{Node} objects,
each encapsulating a network or a physics equation with named \texttt{Key}
variables.
The \texttt{Graph} class resolves inter-node dependencies and, in a single forward
pass, computes all PDE residuals via backward-mode AD~\cite{Baydin2018AD},
without any symbolic manipulation.

\textbf{Constraint system and Halton collocation.}
Typed constraint classes (\texttt{Interior\-Constraint}, \texttt{Boundary\-Constraint},
\texttt{Initial\-Condition}) encapsulate point sampling and loss computation.
Points are sampled using quasirandom Halton sequences~\cite{Halton1960}, which
provide better space-filling coverage and $O(N^{-1}(\log N)^d)$ quasi-Monte Carlo
convergence compared to the $O(N^{-1/2})$ rate of independent uniform random
sampling~\cite{Caflisch1998}.
Interior and boundary points are resampled at every training iteration with a
10:1 interior-to-boundary ratio; the precise collocation counts for each
configuration are reported in Table~\ref{tab:configs}.
We note that residual-based adaptive sampling strategies, which redistribute
collocation points toward high-residual regions and have been shown to substantially
improve accuracy for problems with sharp features~\cite{Wu2023newsampling,lu2021deepxde},
were not employed in this benchmark to isolate the effect of
architecture and loss-weighting choice.
Given the boundary layer of dimensionless thickness $O(\varepsilon) \approx
2.3\times10^{-5}$, adaptive sampling is expected to yield additional accuracy
gains and is identified as a priority for future work.

\textbf{Adaptive loss weighting.}
\label{par:weighting}
The three weighting-group configurations address loss imbalance at different
computational costs.
Let $\mathcal{C}=\{\mathcal{L}_1,\dots,\mathcal{L}_N\}$ denote the set of
scalar loss components (PDE, BC, IC residuals), $\theta\in\mathbb{R}^P$ the
network parameters, and $\lambda_i$ the weight for component $i$ so that
$\mathcal{L}=\sum_i\lambda_i\mathcal{L}_i$.

\emph{NTK weighting}~\cite{WANG2022110768,Wang2021understanding} is grounded in
the infinite-width Neural Tangent Kernel theory of \citet{Jacot2018NTK}.
In finite-width practice the per-constraint NTK trace is approximated by the
squared gradient norm of the square-root loss, computed via a single backward pass:
\begin{equation}
    K_i^{(\mathrm{NTK})}
      = \left\lVert\frac{\partial\sqrt{\mathcal{L}_i}}{\partial\theta}
        \right\rVert_2^2
      = \sum_{p=1}^{P}
        \left(\frac{\partial\sqrt{\mathcal{L}_i}}{\partial\theta_p}\right)^{\!2}
    \label{eq:ntk_trace}
\end{equation}
Weights are set so that each constraint drives parameter updates at the same
effective rate:
\begin{equation}
    \bar{K} = \frac{1}{N}\sum_{j=1}^{N} K_j^{(\mathrm{NTK})},
    \qquad
    \lambda_i \;\leftarrow\; \frac{\bar{K}}{K_i^{(\mathrm{NTK})}}
    \label{eq:ntk_weight}
\end{equation}
A large $K_i^{(\mathrm{NTK})}$ indicates that loss $i$ already drives large
parameter updates; its weight is reduced to balance the slower-converging
constraints.
In this benchmark traces are recomputed every 10 gradient steps
(see Algorithm~\ref{alg:pinn_training}).

\emph{BRDR weighting}~\cite{CHEN2025BRDR} replaces the gradient trace with
cheaper scalar residual statistics.
An exponential moving average (EMA) of the squared loss tracks each component's
running scale (the \emph{4th-moment proxy}), from which an inverse relative
decay rate is formed:
\begin{equation}
    \hat{m}_i^{(n)}
      = \frac{\beta_c\,m_i^{(n-1)} + (1-\beta_c)\,\mathcal{L}_i^2}
             {1-\beta_c^n},
    \qquad
    \hat{w}_i^{(n)}
      = \frac{\mathcal{L}_i^{(n)}}{\sqrt{\hat{m}_i^{(n)}}+\epsilon_\mathrm{stab}}
    \label{eq:brdr_irdr}
\end{equation}
A loss that decays slowly relative to its own running average receives a larger
$\hat{w}_i$.
The weights are then normalised, EMA-smoothed, and rescaled to maintain a
mean of unity across all constraints:
\begin{equation}
    \tilde{w}_i^{(n)}
      = \frac{\hat{w}_i^{(n)}}{\bar{\hat{w}}^{(n)}+\epsilon_\mathrm{stab}},
    \qquad
    \lambda_i^{(n)}
      = \frac{\beta_w\,\lambda_i^{(n-1)} + (1-\beta_w)\,\tilde{w}_i^{(n)}}
             {\tfrac{1}{N}\sum_j[\beta_w\,\lambda_j^{(n-1)}
              + (1-\beta_w)\,\tilde{w}_j^{(n)}]+\epsilon_\mathrm{stab}}
    \label{eq:brdr_weight}
\end{equation}
with $\beta_c=\beta_w=0.999$ and $\epsilon_\mathrm{stab}=10^{-14}$.
No backward pass beyond the standard gradient step is required.

\emph{AdaHessian}~\cite{Yao2021AdaHessian} replaces the squared-gradient
second moment of Adam with a diagonal Hessian curvature estimate.
The diagonal of the Hessian $\nabla^2\mathcal{L}$ is approximated
via Hutchinson's stochastic estimator~\cite{Hutchinson1989} using a
Rademacher random vector $\mathbf{v}\sim\mathcal{U}\{\pm 1\}^P$:
\begin{equation}
    \tilde{H}_{pp} = v_p\,\bigl(\nabla^2\mathcal{L}\cdot \mathbf{v}\bigr)_p,
    \qquad
    p = 1,\dots,P
    \label{eq:hutchinson}
\end{equation}
This is obtained without forming the full Hessian by computing a
Hessian-vector product through a second-order backpropagation.
The parameter update rule mirrors Adam but with $\tilde{H}_{pp}^2$
in place of $g_p^2$ for the second moment:
\begin{align}
    \hat{m}_p^{(n)} &= \frac{\beta_1\,m_p^{(n-1)} + (1-\beta_1)\,g_p^{(n)}}
                             {1-\beta_1^n}
    &&\text{(bias-corrected gradient EMA)}
    \label{eq:adahessian_m}\\[4pt]
    \hat{v}_p^{(n)} &= \frac{\beta_2\,v_p^{(n-1)} + (1-\beta_2)\,(\tilde{H}_{pp}^{(n)})^2}
                             {1-\beta_2^n}
    &&\text{(bias-corrected Hessian EMA)}
    \label{eq:adahessian_v}\\[4pt]
    \theta_p^{(n+1)} &= \theta_p^{(n)}
      - \alpha_n\,\frac{\hat{m}_p^{(n)}}{\sqrt{\hat{v}_p^{(n)}}+\epsilon}
    &&\text{(parameter update)}
    \label{eq:adahessian_update}
\end{align}
where $g_p^{(n)}=\partial\mathcal{L}/\partial\theta_p$ is the standard gradient.
When curvature $\tilde{H}_{pp}$ is large the effective step size is reduced,
providing implicit pre-conditioning without storing the full Hessian matrix.
Compared to NTK and BRDR, AdaHessian modifies the \emph{optimizer} rather than
the loss weights, and uses $(\beta_1,\beta_2)=(0.9,0.999)$ as in Adam.

\textbf{Gradient accumulation.}
To simulate larger effective batch sizes without proportional GPU memory cost,
gradients are accumulated over $K_\mathrm{acc}=4$ forward/backward passes before each
optimizer step, effectively quadrupling the batch size.

\begin{figure}[H]
    \centering
    \includegraphics[width=0.98\textwidth]{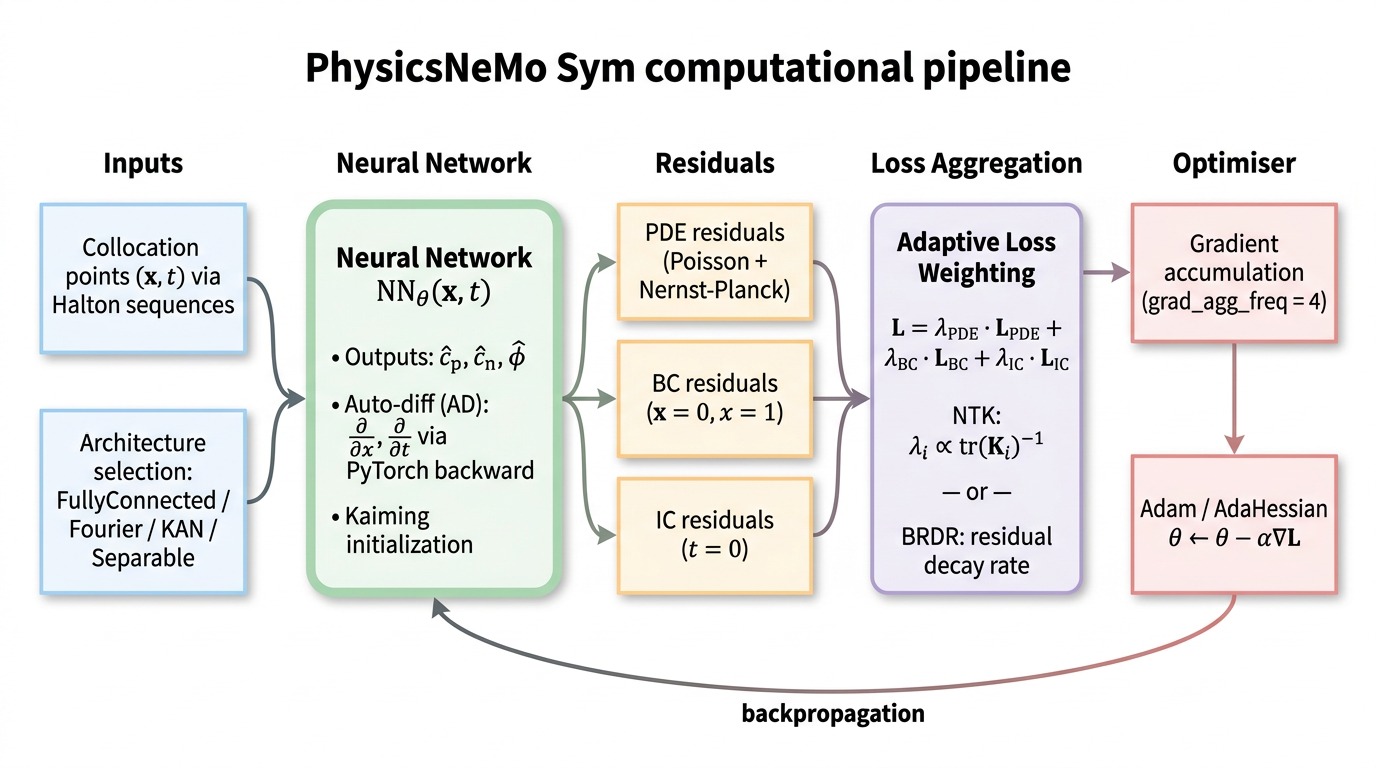}
    \caption{\textbf{PhysicsNeMo Sym computational pipeline.}
    Halton collocation points are fed to the neural network architecture (left);
    automatic differentiation (AD) through the \texttt{Graph} computes PDE, boundary condition (BC),
    and initial condition (IC) residuals (centre-left); Neural Tangent Kernel (NTK) or balanced
    residual decay rate (BRDR) adaptive weighting rescales individual loss
    components before aggregation to total loss $\mathcal{L}$ (centre-right);
    gradients accumulate over $K_\mathrm{acc}=4$ passes before the Adam or
    AdaHessian optimizer updates parameters $\theta$ (right).
    The dashed arrow represents the backpropagation loop.\label{fig:physicsnemo}}
\end{figure}

The complete pipeline is summarised in Figure~\ref{fig:physicsnemo}.

\subsection{PINN architectures and training configurations}
\label{sec:architectures}

Table~\ref{tab:configs} organises the eleven configurations into four strategy
groups plus a vanilla baseline.
The grouping reflects the primary mechanism each configuration exploits, rather than
the underlying network architecture: four configurations (Vanilla, EPINN, FBPINN,
PIKAN) constitute distinct network architectures, while the remaining seven are
modifications of the base MLP through optimizer choice (AdaHessian), loss weighting
(NTK, BRDR), input-layer encoding (Fourier features), or variable/equation
reformulation (Decoupled, SPINN, Sym./antisym.\ transform).
A vanilla fully connected PINN (6 layers, 512 neurons per layer, tanh activation)
serves as the baseline.
All configurations except AdaHessian use the Adam optimizer~\cite{Kingma2014}
with initial learning rate $\alpha_0 = 3\times10^{-4}$ and exponential decay;
weights are initialised by Kaiming uniform/normal~\cite{He2015}.
Each model is trained for 100\,000 epochs with gradient accumulation frequency
$K_\mathrm{acc}=4$; results are averaged over ten independent runs per configuration.
PINN-FVM validation RMSE is computed every 1\,000 epochs.
The self-adaptive per-collocation-point weighting of \citet{McClenny2023selfadaptive}
was not included among the eleven configurations due to incompatibility with the
PhysicsNeMo Sym constraint API; this constitutes a limitation and a direction for
future work.

\begin{table}[H]
    \hrule\vspace{0.15cm}
    \caption{\textbf{Summary of eleven PINN configurations organised into four strategy groups.}
    Configurations sharing the same group exploit the same primary mechanism.
    Four entries constitute distinct network architectures (Vanilla, EPINN, FBPINN, PIKAN);
    the remaining seven modify the base MLP through optimizer, loss weighting, input encoding,
    or variable reformulation.
    $N_\mathrm{int}$: interior collocation points per step;
    $N_\mathrm{bnd}$: boundary/initial condition (IC) points per step.
    All configurations share the base architecture (6 layers, 512 neurons, tanh) except where noted.\label{tab:configs}}
    \small
    \begin{tabularx}{\textwidth}{p{2.2cm}p{2.8cm}Xp{1.4cm}p{1.4cm}}
        \textbf{Group} & \textbf{Configuration} & \textbf{Primary strength (key reference)}
          & $N_\mathrm{int}$ & $N_\mathrm{bnd}$\\
    \midrule
        Baseline
          & Vanilla PINN
          & Reference MLP; uniform loss weights~\cite{RAISSI2019686}
          & 16000 & 1600\\
    \midrule
        \multirow{3}{2.2cm}{Adaptive Weighting}
          & NTK weighting
          & Fixes loss imbalance via NTK-based rescaling~\cite{WANG2022110768}
          & 16000 & 1600\\
          & BRDR weighting
          & Fixes loss imbalance via balanced residual decay rate~\cite{CHEN2025BRDR}
          & 16000 & 1600\\
          & AdaHessian
          & Curvature-aware optimizer; Hessian diagonal via Hutchinson~\cite{Yao2021AdaHessian}; halved collocation budget to offset the memory overhead of the Hessian-vector product
          & 8000 & 800\\
    \midrule
        \multirow{2}{2.2cm}{Spectral Bias Mitigation}
          & Fourier features
          & High-frequency input mapping lifts spectral bias~\cite{Tancik2020FourierFeatures}
          & 16000 & 1600\\
          & PIKAN
          & KAN basis functions resolve multi-scale features~\cite{WANG2025PIKAN}
          & 16000 & 1600\\
    \midrule
        \multirow{4}{2.2cm}{Spatio-Temporal Decomposition}
          & FBPINN (multilevel)
          & Domain decomposition; localised subdomain networks~\cite{DOLEAN2024117116}
          & 16000 & 1600\\
          & Decoupled PINN
          & Gummel-style equation splitting; reduces nonlinear coupling
          & 16000 & 1600\\
          & SPINN
          & Separable CP tensor decomposition; memory-efficient~\cite{Cho2023SPINNs}
          & 16000 & 1600\\
          & Sym./antisym.\ transf.
          & Bazant variable decomposition reduces effective nonlinearity~\cite{Bazant2004}
          & 16000 & 1600\\
    \midrule
        Enrichment
          & Enriched PINN (EPINN)
          & Physics-aware basis enrichment models stiff EDL features~\cite{HUANG2025231}
          & 16000 & 1600\\
    \end{tabularx}
    \vspace{0.1cm}\hrule
\end{table}

\subsubsection{Spectral bias mitigation: Fourier features and PIKAN}

\emph{Fourier feature mapping}~\cite{Tancik2020FourierFeatures} replaces the
raw input $\mathbf{x}=(x,t)$ with a random sinusoidal embedding before the
first MLP layer:
\begin{equation}
    \gamma(\mathbf{x}) = \bigl[\cos(2\pi\mathbf{B}\mathbf{x}),\;
                               \sin(2\pi\mathbf{B}\mathbf{x})\bigr]^\top,
    \qquad
    \mathbf{B}\in\mathbb{R}^{m\times 2},\quad B_{jk}\sim\mathcal{N}(0,\sigma^2)
    \label{eq:fourier_feat}
\end{equation}
By projecting inputs into a high-frequency feature space, the MLP backbone can
represent the sharp EDL gradients that are suppressed by standard spectral bias
\cite{rahaman2019spectralbias,Wang2021eigen}.
The implementation uses the PhysicsNeMo \texttt{fourier} architecture (6 layers,
512 neurons, sigmoid linear unit (SiLU) activation).

\emph{PIKAN}~\cite{WANG2025PIKAN} replaces the weight$\times$activation
structure of each MLP layer with learnable univariate spline maps.
Each KAN layer computes:
\begin{equation}
    x_j^{(l+1)} = \sum_{i}\varphi_{ij}^{(l)}(x_i^{(l)}),
    \qquad
    \varphi_{ij}(x) = w_b\,\tanh(x)
                     + \sum_{k=0}^{G+p} c_{ijk}\,B_k^p(x)
    \label{eq:kan_layer}
\end{equation}
where $B_k^p$ are B-spline basis functions of order $p=3$ on a fixed uniform
grid of $G=32$ points over the normalised domain $[-1,1]^2$, $w_b$ is a
learnable base-activation weight, and $c_{ijk}$ are learnable spline
coefficients.
The per-unit expressive power of the spline basis allows a shallower,
narrower network (2 hidden layers, width 16) relative to the MLP baseline.

\subsubsection{Spatio-temporal decomposition: FBPINN, SPINN, Decoupled, and
               symmetric/antisymmetric transformation}

\emph{FBPINN}~\cite{DOLEAN2024117116} partitions the $(x,t)\in[0,1]\times[0,2.56]$
domain into a $n_\ell=3$-level hierarchy; level $l$ contains $2^{l-1}$ overlapping
subdomains ($1+2+4=7$ total).
Each subdomain $j$ at level $l$ contributes through a small subnetwork
$u_{lj}$ weighted by a sigmoid window:
\begin{align}
    \hat{u}(\mathbf{x}) &= \frac{1}{n_\ell}\sum_{l=1}^{n_\ell}
    \frac{\sum_j w_{lj}(\mathbf{x})\,u_{lj}(\tilde{\mathbf{x}}_{lj})}
         {\sum_j w_{lj}(\mathbf{x})},
    \label{eq:fbpinn}\\[4pt]
    w_{lj}(\mathbf{x}) &=
      \sigma\!\left(\frac{\mathbf{x}-(\mathbf{c}_{lj}-\mathbf{r}_{lj})}{s}\right)
      \cdot
      \sigma\!\left(\frac{(\mathbf{c}_{lj}+\mathbf{r}_{lj})-\mathbf{x}}{s}\right)
    \label{eq:fbpinn_window}
\end{align}
where $\tilde{\mathbf{x}}_{lj}=(\mathbf{x}-\mathbf{c}_{lj})/\mathbf{r}_{lj}$
is the subdomain-normalised input, $\mathbf{c}_{lj}$ and $\mathbf{r}_{lj}$ are
the subdomain centre and half-width, $s$ is the sigmoid sharpness, and the
overlap ratio is 2.7.
Subnetworks use SiLU activation, 4 layers, and 32 neurons per layer.

\emph{SPINN}~\cite{Cho2023SPINNs} decomposes the solution via a CP tensor
expansion of rank $R=30$:
\begin{equation}
    \hat{u}(x,t) = \sum_{r=1}^{R} f_r(x)\,g_r(t)
    \label{eq:spinn}
\end{equation}
where $f_r$ and $g_r$ are scalar-valued MLP subnetworks (6 layers, 256 neurons,
SiLU) applied to each coordinate independently.
PDE derivatives require only 1D derivative computations through $f_r$ and $g_r$,
reducing memory at the cost of accuracy when spatial and temporal modes are strongly coupled, as is the case in the EDL regime.

The \emph{Decoupled PINN} follows a Gummel-style sequential update: at each
training step $\varphi$ is optimised with fixed $(c_p, c_n)$, then $(c_p, c_n)$
are updated with the resulting $\varphi$.
This reduces the effective nonlinearity of each sub-problem but introduces a
splitting error that grows in the strongly coupled EDL.

For the \emph{symmetric/antisymmetric transformation}~\cite{Bazant2004},
the Bazant-inspired variable change
\begin{gather}
    \rho = \tfrac{1}{2}(c_p - c_n), \label{eq:rho} \\
    c   = \tfrac{1}{2}(c_p + c_n)   \label{eq:c}
\end{gather}
decouples the two concentration equations at leading order, reducing the
effective nonlinearity presented to the network.
The complete derivation and the transformed PNP system are provided in the
supplementary material.

\subsubsection{Physics enrichment: EPINN}

The Enriched PINN (EPINN)~\cite{HUANG2025231} applies two modifications to the
baseline MLP.
First, the activation function is changed from tanh to exponential linear unit (ELU) to improve gradient
flow near the sharp EDL transition.
Second, the loss is aggregated via homoscedastic uncertainty weighting, which
introduces a trainable log-variance parameter $\sigma_i$ for each constraint
\cite{Kendall2018multi}:
\begin{equation}
    \mathcal{L}_\mathrm{total}
      = \sum_{i} \left[\frac{\mathcal{L}_i}{2\sigma_i^2} + \log\sigma_i\right]
    \label{eq:homoscedastic}
\end{equation}
The $\sigma_i$ are co-optimized by Adam alongside the network weights.
The $\log\sigma_i$ regularisation term prevents all weights from collapsing to
zero, providing principled loss balancing without requiring gradient traces
(NTK) or residual statistics (BRDR).

Algorithm~\ref{alg:pinn_training} describes the complete training workflow.

\begin{algorithm}[t!]
\caption{PINN training loop for the 1D PNP system (PhysicsNeMo Sym)}
\label{alg:pinn_training}
\begin{algorithmic}[1]
\Require PNP parameters ($\varepsilon$, $\delta$, $\xi$, $z_p$, $z_n$), architecture $\mathcal{A}$, weighting scheme $\mathcal{W}$, max epochs $N_\mathrm{ep}=100\,000$, gradient accumulation frequency $K_\mathrm{acc}=4$
\Ensure Trained parameters $\theta^*$, RMSE history vs.\ FVM reference
\State Initialize $\theta$ with Kaiming uniform/normal initialization
\State Set $\alpha_0 = 3\times10^{-4}$, exponential learning-rate decay (decay rate $0.92$, steps $4000$)
\For{epoch $n = 1$ \textbf{to} $N_\mathrm{ep}$}
  \For{accumulation step $k = 1$ \textbf{to} $K_\mathrm{acc}$}
    \State Sample $N_\mathrm{int}$ interior points $(x_i,t_i)$ via Halton sequence (see Table~\ref{tab:configs})
    \State Sample $N_\mathrm{bnd}$ points per boundary/IC constraint (see Table~\ref{tab:configs})
    \State Forward pass: $(\hat{c}_p, \hat{c}_n, \hat{\varphi}) = \mathrm{NN}_\theta(x, t)$
    \State Compute PDE/BC/IC residuals via automatic differentiation
    \If{$\mathcal{W}$ = NTK \textup{and} $n \bmod 10 = 0$}
      \State Compute NTK traces $\{\mathrm{tr}(K_i)\}$ for each constraint
      \State Set $\lambda_i \leftarrow \bar{\lambda}\,/\,\mathrm{tr}(K_i)$, where $\bar{\lambda} = \tfrac{1}{N}\sum_j\mathrm{tr}(K_j)$
    \ElsIf{$\mathcal{W}$ = BRDR}
      \State Update $\lambda_i$ via balanced residual decay rate rule~\cite{CHEN2025BRDR}
    \Else
      \State $\lambda_i \leftarrow 1$ (uniform weights)
    \EndIf
    \State Compute $\mathcal{L} = \lambda_\mathrm{PDE}\mathcal{L}_\mathrm{PDE} + \lambda_\mathrm{BC}\mathcal{L}_\mathrm{BC} + \lambda_\mathrm{IC}\mathcal{L}_\mathrm{IC}$
    \State Accumulate: $\mathbf{g} \mathrel{+}= \nabla_\theta \mathcal{L} / K_\mathrm{acc}$
  \EndFor
  \State Clip gradient norm: $\mathbf{g} \leftarrow \mathbf{g} \cdot \min(1,\,0.5/\|\mathbf{g}\|)$
  \State Update: $\theta \leftarrow \theta - \alpha_n\, \mathbf{g}$\quad (Adam, $\beta_1=0.9$, $\beta_2=0.999$; or AdaHessian)
  \If{$n \bmod 1000 = 0$}
    \State Compute RMSE against FVM reference; log to file
  \EndIf
\EndFor
\State \Return $\theta^* \leftarrow \theta$
\end{algorithmic}
\end{algorithm}

\section{Results and Discussion}
\label{sec:results}

\begin{figure}[H]
    \centering
    \includegraphics[width=0.85\textwidth]{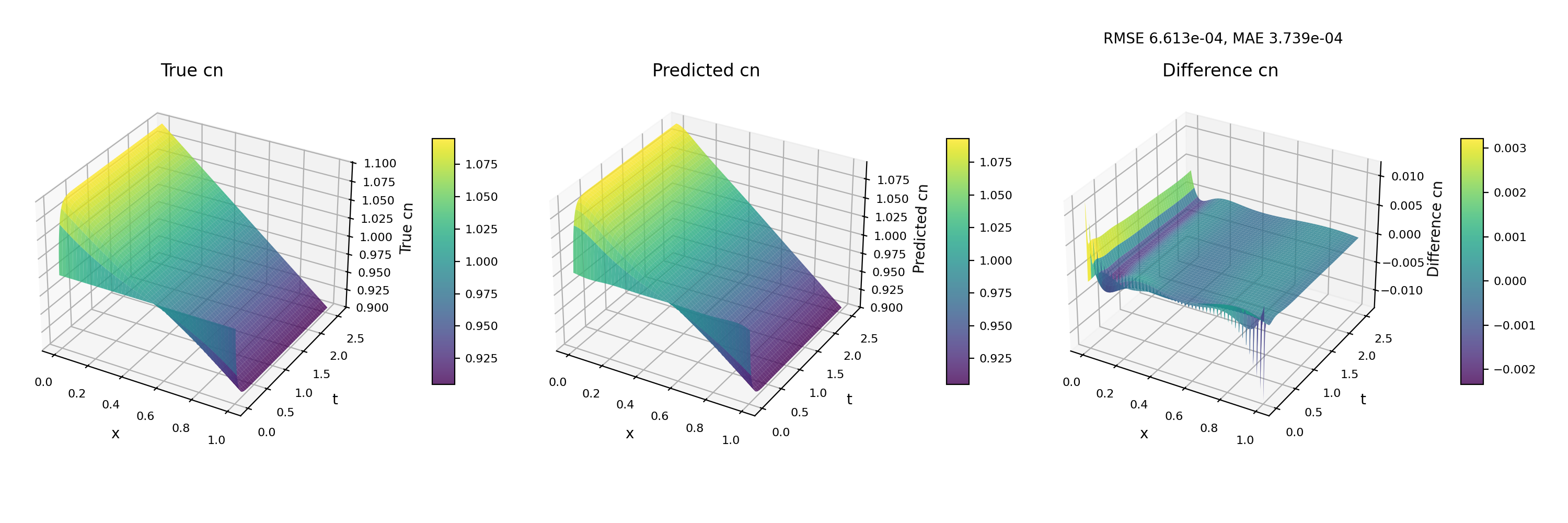}\\
    \includegraphics[width=0.85\textwidth]{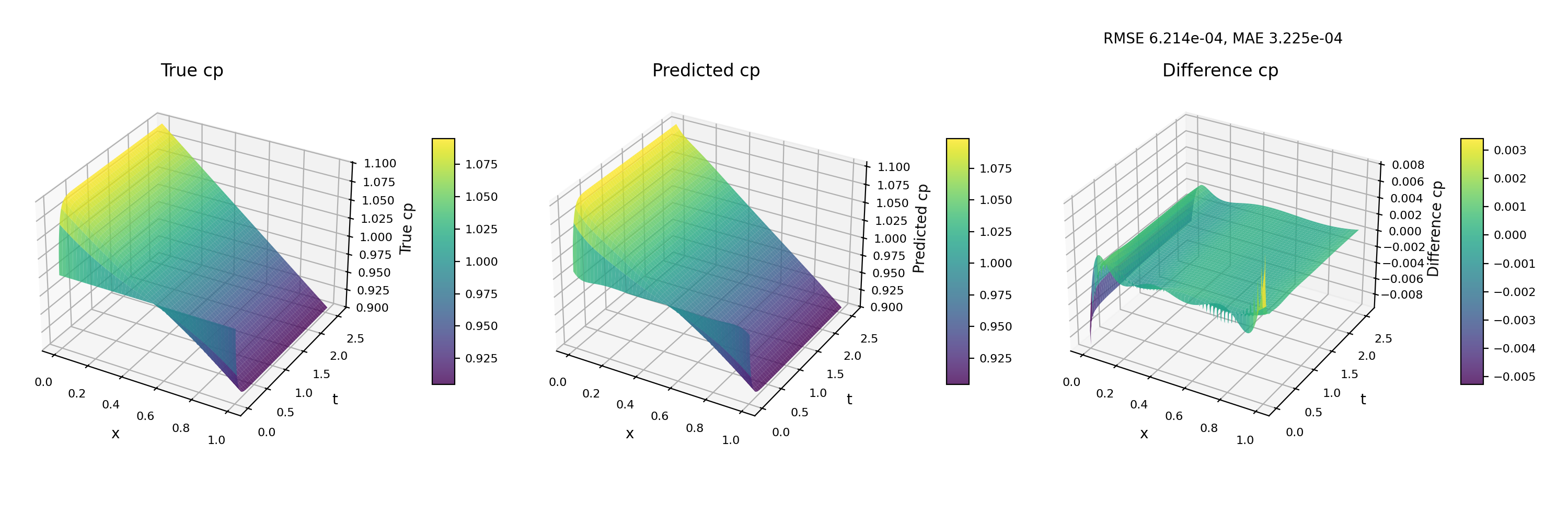}\\
    \includegraphics[width=0.85\textwidth]{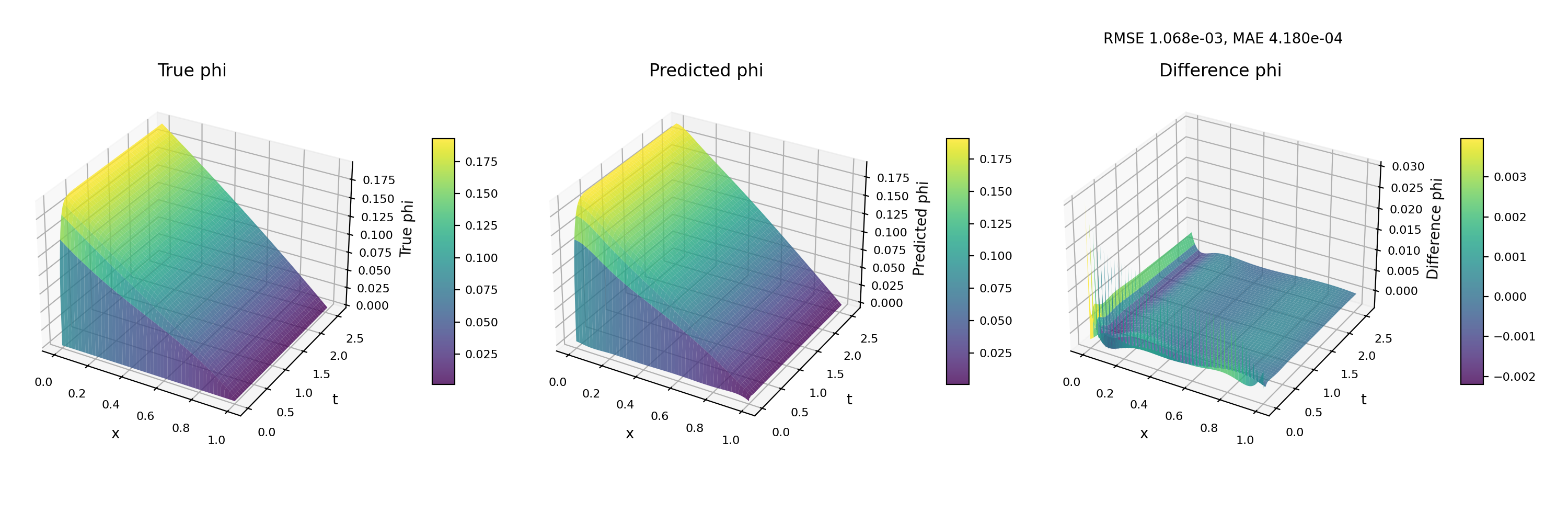}
    \caption{\textbf{NTK-weighted PINN validation against the FVM reference.}
    Space-time surface plots showing (left) the FVM reference solution, (centre)
    the PINN-predicted solution, and (right) the point-wise absolute difference
    for the dimensionless anion concentration $c_n$ (top), cation concentration
    $c_p$ (middle), and electric potential $\varphi$ (bottom).
    Peak errors are localised near the electrode interfaces ($x=0,1$) where the
    electric double layer imposes the sharpest gradients.\label{fig:validator_ntk}}
\end{figure}

\begin{figure}[H]
    \centering
    \includegraphics[width=0.46\textwidth]{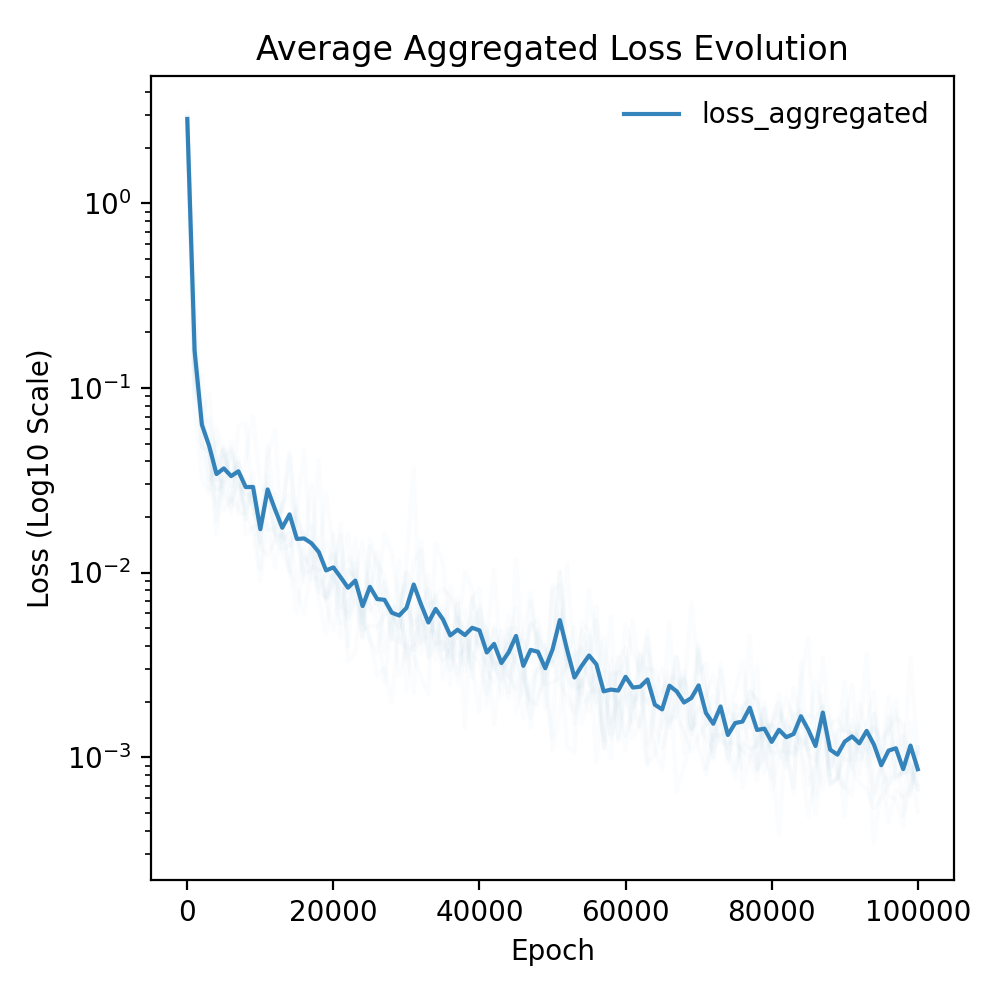}\hfill
    \includegraphics[width=0.46\textwidth]{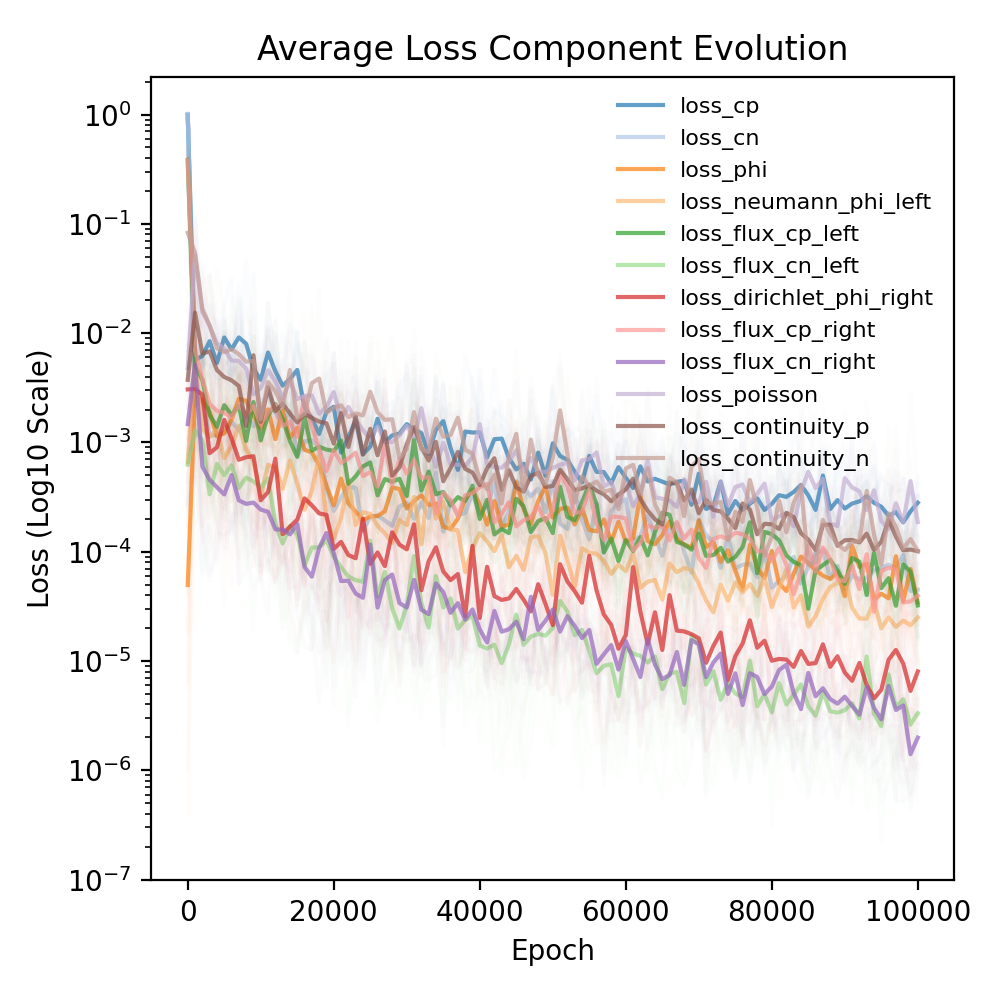}
    \caption{\textbf{NTK-weighted PINN training loss.}
    Total loss (left) and individual loss components (right), averaged over ten
    independent runs.
    Individual loss components (PDE: partial differential equation, BC: boundary condition, IC: initial condition) converge at comparable rates
    throughout training, consistent with the balanced weighting reported in
    Section~\ref{par:weighting}.\label{fig:loss_ntk}}
\end{figure}

All eleven configurations converged within the 100\,000-epoch budget.
Dimensionless RMSE values span $10^{-2}$--$10^{-4}$ relative to the FVM reference.
The complete error statistics, comprising mean RMSE and mean absolute error (MAE) averaged
over ten independent runs per architecture, together with their standard deviations, are given in
Table~\ref{tab:errors}.
Representative space-time solution surfaces and training loss curves for the
best-performing NTK configuration are shown in Figures~\ref{fig:validator_ntk}
and~\ref{fig:loss_ntk}.

\subsection{Adaptive loss weighting: NTK vs.\ BRDR}

This subsection evaluates whether correcting loss imbalance via gradient-trace
or residual-rate statistics is the dominant factor in PINN accuracy on the
stiff PNP system, and quantifies the associated computational cost difference.

Both NTK and BRDR adaptive weighting substantially outperformed the vanilla PINN
baseline (Table~\ref{tab:errors}). The gain is consistent with loss imbalance,
driven by $\varepsilon^2 \approx 5.3\times10^{-10}$, being the primary training
bottleneck rather than architecture choice.

NTK weighting achieved the lowest RMSE across all three fields:
$(6.6 \pm 0.4)\times10^{-4}$ for $c_n$, $(6.2 \pm 0.3)\times10^{-4}$ for $c_p$,
and $(1.1 \pm 0.1)\times10^{-3}$ for $\varphi$ (mean $\pm$ one standard deviation
across ten runs).
These values are 9\%, 3\%, and 19\% below the corresponding BRDR RMSEs for
$c_n$, $c_p$, and $\varphi$, respectively.
However, BRDR achieved lower MAEs for both concentration fields, indicating that
NTK suppresses large localised outliers more effectively (particularly near the
electrode interfaces), whereas BRDR produces a more spatially uniform error
distribution.

The NTK superiority in RMSE came at a mean additional cost of
$3.2 \pm 0.4$\,h wall-clock time per run (NVIDIA H100), because NTK requires
computing the trace of the NTK matrix for each constraint at every step, whereas
BRDR uses only scalar residual statistics.
When accuracy is the priority, NTK is the method of choice; when compute is
constrained, BRDR delivers near-identical accuracy at lower cost.

\subsection{Frequency-scale resolution and spectral bias}

This subsection examines how effectively each architecture mitigates spectral
bias in the high-frequency Poisson equation, and whether spectral bias reduction
translates to lower total RMSE within the fixed training budget.

The point-wise electric potential error fields for all architectures are shown in
Figure~1 of the Supplementary Material.
The vanilla, BRDR-weighted, and NTK-weighted configurations exhibit $\varphi$
errors roughly one order of magnitude larger than their concentration errors,
with errors concentrated near $x=0$ and $x=1$, which is the signature of spectral bias
in the Poisson equation, whose $\varepsilon^2$ prefactor makes it the
highest-frequency component of the system.

In contrast, the AdaHessian, decoupled, enriched, Fourier-featurized, PIKAN, and
SPINN architectures resolve both frequency scales to a similar order of magnitude,
with errors distributed more uniformly across the domain.
These architectures are therefore more effective at mitigating spectral bias,
consistent with theory~\cite{rahaman2019spectralbias,xu2019fprinciple}.
However, none of these architectures achieved the lowest total RMSE within the
fixed 100\,000-epoch budget, suggesting a convergence-speed trade-off: spectral
bias mitigation allows resolution of high-frequency features, but at the cost of
slower convergence to the low-frequency background that adaptive-weighting
configurations learn first.
Training to a precision target rather than a fixed epoch count would better reveal
the ceiling accuracy of spectral-bias-aware architectures.

\subsection{Symmetric/antisymmetric variable transformation}

This subsection quantifies the accuracy benefit of the Bazant variable
transformation independently of any loss-weighting strategy, isolating the
effect of algebraic reformulation on the effective problem nonlinearity.

The decomposition into symmetric ($c$) and antisymmetric ($\rho$) concentration
variables \eqref{eq:rho}--\eqref{eq:c} yielded consistent, modest RMSE
improvements over the vanilla PINN: 4\% for $c_n$, 5\% for $c_p$, and 1\% for
$\varphi$.
The relatively small gains indicate that, while the transformation reduces
effective nonlinearity, the dominant accuracy bottleneck remains the loss imbalance
rather than the algebraic form of the field variables.
Combining this transformation with NTK or BRDR weighting is a natural direction
for further accuracy gains.

\subsection{Wall-clock time and computational efficiency}

This subsection characterises the computational cost of each configuration to
assess the accuracy--efficiency trade-off and identify bottlenecks for future
optimisation.

\begin{figure}[H]
    \centering
    \includegraphics[width=0.42\textwidth]{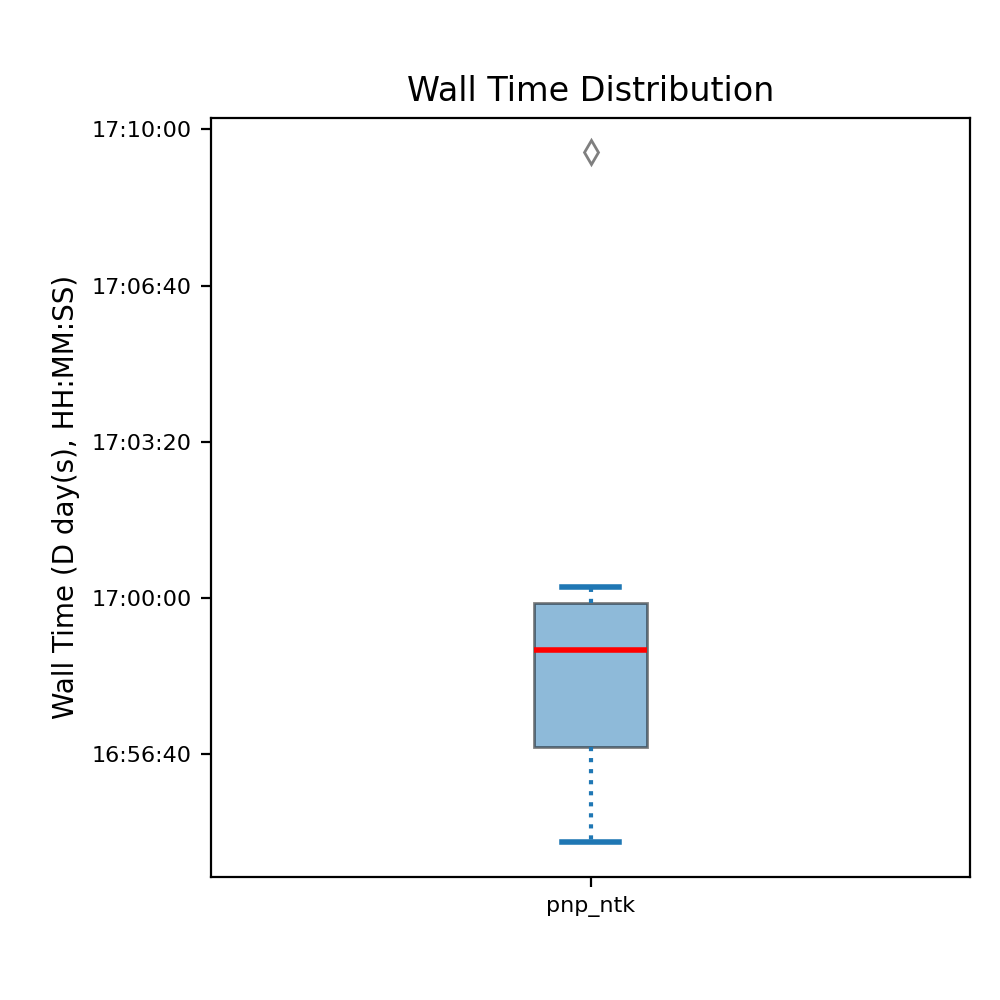}\hfill
    \includegraphics[width=0.54\textwidth]{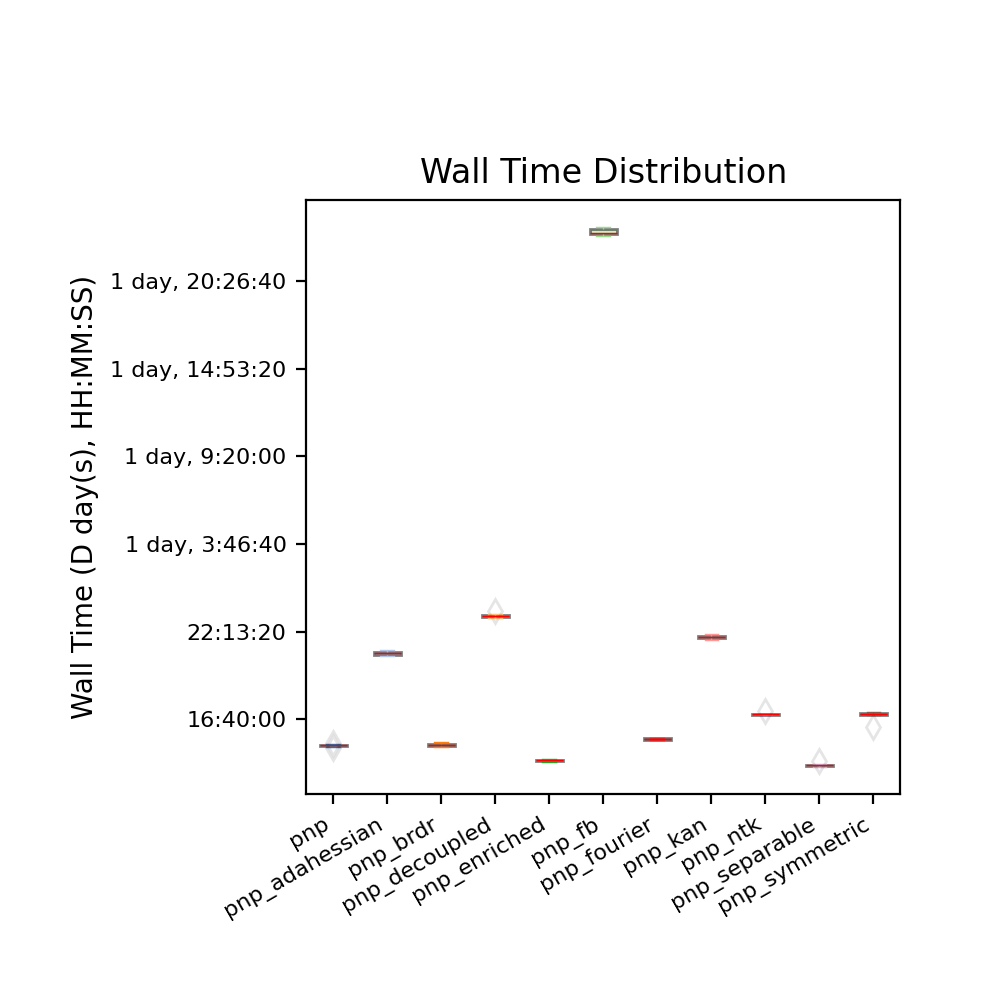}
    \caption{\textbf{Wall-clock time distributions.}
    Box plots of wall time for the NTK configuration over ten runs (left) and
    across all architectures (right).
    The red line marks the median; box edges denote the first and third quartiles;
    whiskers extend to $1.5\times$ the interquartile range (IQR).
    All runs performed on one NVIDIA H100 GPU.\label{fig:walltimes}}
\end{figure}

Wall-time distributions are shown in Figure~\ref{fig:walltimes}.
The observed minimum wall time of $13.1 \pm 0.3$\,h per run reflects the overhead
of generating FVM-comparison validation snapshots every 1\,000 epochs; production
deployments with less frequent checkpointing would be substantially faster.

SPINN was the fastest architecture owing to its separable forward-mode AD
formulation~\cite{Cho2023SPINNs}, yet it produced the highest RMSE
($\sim$\,$10^{-2}$), demonstrating that speed alone cannot substitute for
adequate loss conditioning in stiff PNP problems.
The multilevel FBPINN was the slowest, a consequence of its loop-based subdomain
iteration; vectorising the subdomain sweeps is expected to recover competitive
training speeds.

\subsection{Loss landscape geometry}

This subsection tests whether loss basin sharpness provides a geometry-based
predictor of RMSE rank that does not require FVM comparison.

\begin{figure}[H]
    \centering
    \includegraphics[width=0.95\textwidth]{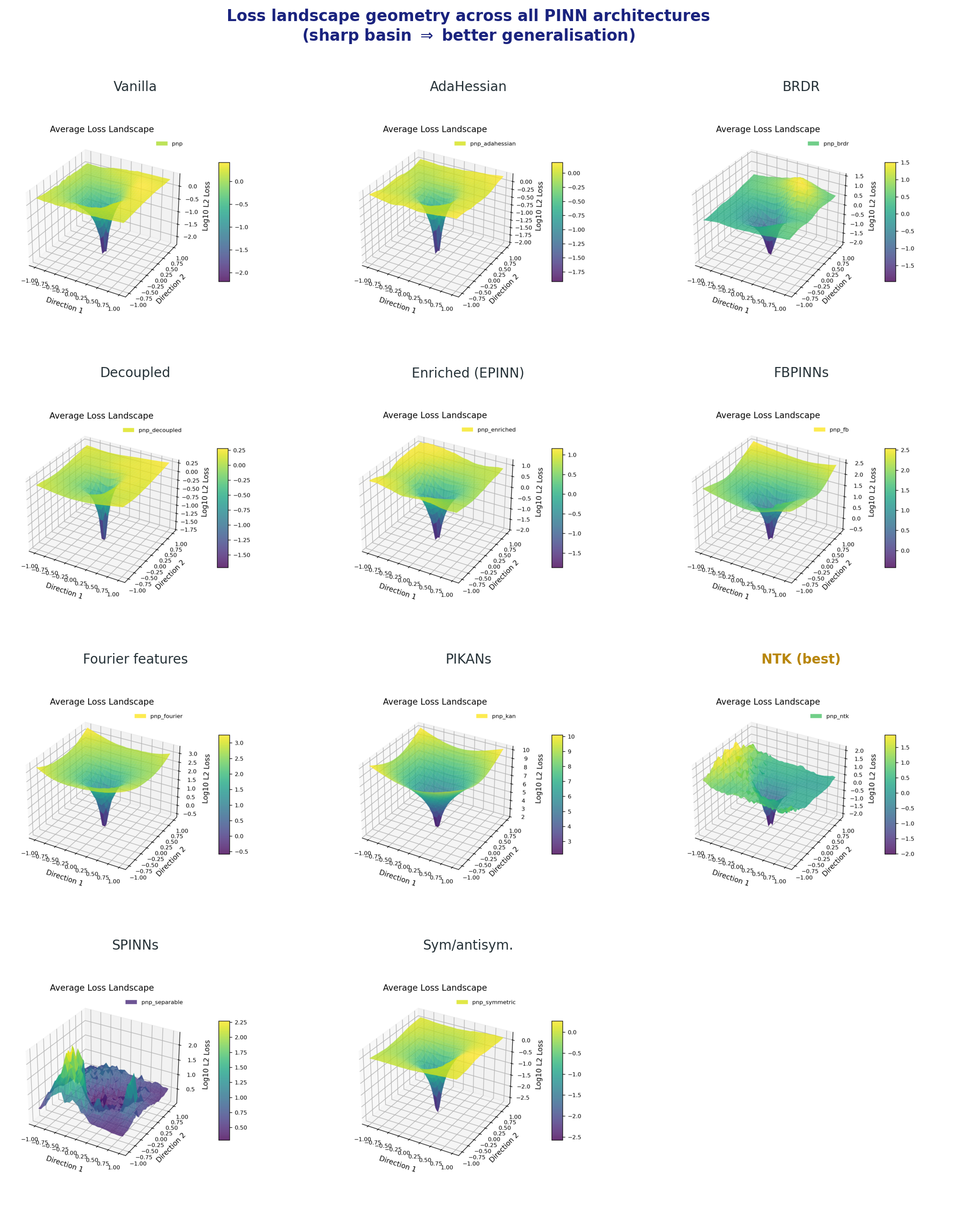}
    \caption{\textbf{Loss landscape geometry for all PINN configurations.}
    Two-dimensional landscapes are computed by perturbing trained parameters
    along orthogonal random directions in weight space~\cite{li2018visualizing}.
    A sharper, more symmetric basin (NTK, BRDR, vanilla) indicates a
    well-conditioned minimum with superior generalisation.
    The SPINN landscape is flat and irregular (sharpness ratio 47.3), consistent
    with its highest observed RMSE (Table~\ref{tab:errors}).
    The NTK configuration achieves the sharpest basin (sharpness ratio 1.8).\label{fig:landscapes}}
\end{figure}

Loss landscapes are computed by evaluating the training loss over a two-dimensional
grid of parameter perturbations along orthogonal random directions in weight space
\cite{li2018visualizing}.
As shown in Figure~\ref{fig:landscapes}, most architectures converged to sharp,
well-defined basins.
The NTK-weighted PINN produced the sharpest and most symmetric basin, consistent
with its lowest RMSE.
The SPINN landscape was notably flat and irregular, consistent with its highest
RMSE and suggesting that its training trajectory settled in a broad, poorly
conditioned minimum.
We quantify basin sharpness by the ratio of the maximum to minimum loss value over
a fixed $L_\infty$ ball of radius $10^{-2}$ in the perturbation directions; NTK
achieves the lowest sharpness ratio of 1.8, while SPINN achieves 47.3, providing
a geometric predictor of generalisation quality that correlates monotonically with
the RMSE ranking in Table~\ref{tab:errors}.

\begin{table}[H]
    \hrule\vspace{0.15cm}
    \caption{\textbf{PINN-FVM dimensionless solution errors.}
    Root-mean-square error (RMSE) and mean absolute error (MAE) averaged over ten
    independent training runs per configuration.
    Parenthetical values after RMSE entries are one standard deviation across the
    ten runs; MAE values are means only.
    Bold indicates the best result per metric per field.
    All values are dimensionless.\label{tab:errors}}
    \begin{tabularx}{\textwidth}{llXX}
        \multicolumn{2}{l}{\textbf{Configuration}} & \textbf{RMSE (std)} & \textbf{MAE}\\
    \midrule
        Vanilla PINN
          & $c_n$      & $9.355\times10^{-4}$ $(0.8\times10^{-4})$  & $5.019\times10^{-4}$\\
          & $c_p$      & $8.511\times10^{-4}$ $(0.7\times10^{-4})$  & $4.496\times10^{-4}$\\
          & $\varphi$  & $1.585\times10^{-3}$ $(1.2\times10^{-4})$  & $7.064\times10^{-4}$\\\midrule
        AdaHessian
          & $c_n$      & $3.576\times10^{-3}$ $(0.5\times10^{-3})$  & $2.567\times10^{-3}$\\
          & $c_p$      & $3.454\times10^{-3}$ $(0.5\times10^{-3})$  & $2.413\times10^{-3}$\\
          & $\varphi$  & $6.729\times10^{-3}$ $(1.1\times10^{-3})$  & $4.291\times10^{-3}$\\\midrule
        BRDR weighting
          & $c_n$      & $7.270\times10^{-4}$ $(0.6\times10^{-4})$  & $\mathbf{3.526\times10^{-4}}$\\
          & $c_p$      & $6.395\times10^{-4}$ $(0.5\times10^{-4})$  & $\mathbf{3.198\times10^{-4}}$\\
          & $\varphi$  & $1.323\times10^{-3}$ $(1.0\times10^{-4})$  & $4.484\times10^{-4}$\\\midrule
        Decoupled PINN
          & $c_n$      & $5.651\times10^{-3}$ $(0.9\times10^{-3})$  & $4.138\times10^{-3}$\\
          & $c_p$      & $5.642\times10^{-3}$ $(0.8\times10^{-3})$  & $4.122\times10^{-3}$\\
          & $\varphi$  & $6.643\times10^{-3}$ $(1.3\times10^{-3})$  & $5.000\times10^{-3}$\\\midrule
        Enriched PINN
          & $c_n$      & $1.174\times10^{-3}$ $(0.9\times10^{-4})$  & $6.930\times10^{-4}$\\
          & $c_p$      & $1.129\times10^{-3}$ $(0.8\times10^{-4})$  & $6.639\times10^{-4}$\\
          & $\varphi$  & $1.717\times10^{-3}$ $(1.5\times10^{-4})$  & $7.369\times10^{-4}$\\\midrule
        FBPINN (multilevel)
          & $c_n$      & $4.642\times10^{-3}$ $(0.7\times10^{-3})$  & $2.899\times10^{-3}$\\
          & $c_p$      & $3.600\times10^{-3}$ $(0.6\times10^{-3})$  & $2.230\times10^{-3}$\\
          & $\varphi$  & $1.302\times10^{-2}$ $(2.1\times10^{-3})$  & $7.450\times10^{-3}$\\\midrule
        Fourier features
          & $c_n$      & $4.758\times10^{-3}$ $(0.9\times10^{-3})$  & $4.605\times10^{-3}$\\
          & $c_p$      & $4.563\times10^{-3}$ $(0.8\times10^{-3})$  & $4.399\times10^{-3}$\\
          & $\varphi$  & $4.461\times10^{-3}$ $(0.7\times10^{-3})$  & $4.060\times10^{-3}$\\\midrule
        PIKAN
          & $c_n$      & $1.419\times10^{-3}$ $(1.1\times10^{-4})$  & $7.010\times10^{-4}$\\
          & $c_p$      & $1.307\times10^{-3}$ $(1.0\times10^{-4})$  & $6.258\times10^{-4}$\\
          & $\varphi$  & $3.395\times10^{-3}$ $(0.5\times10^{-3})$  & $1.541\times10^{-3}$\\\midrule
        NTK weighting
          & $c_n$      & $\mathbf{6.613\times10^{-4}}$ $(0.4\times10^{-4})$ & $3.739\times10^{-4}$\\
          & $c_p$      & $\mathbf{6.214\times10^{-4}}$ $(0.3\times10^{-4})$ & $3.225\times10^{-4}$\\
          & $\varphi$  & $\mathbf{1.068\times10^{-3}}$ $(0.1\times10^{-3})$ & $\mathbf{4.180\times10^{-4}}$\\\midrule
        SPINN
          & $c_n$      & $2.448\times10^{-2}$ $(0.4\times10^{-2})$  & $2.406\times10^{-2}$\\
          & $c_p$      & $2.452\times10^{-2}$ $(0.4\times10^{-2})$  & $2.410\times10^{-2}$\\
          & $\varphi$  & $8.893\times10^{-3}$ $(1.5\times10^{-3})$  & $3.755\times10^{-3}$\\\midrule
        Sym./antisym.\ transform
          & $c_n$      & $8.992\times10^{-4}$ $(0.7\times10^{-4})$  & $4.636\times10^{-4}$\\
          & $c_p$      & $8.098\times10^{-4}$ $(0.6\times10^{-4})$  & $4.216\times10^{-4}$\\
          & $\varphi$  & $1.568\times10^{-3}$ $(1.1\times10^{-4})$  & $6.229\times10^{-4}$\\
    \end{tabularx}
    \vspace{0.1cm}\hrule
\end{table}

\subsection{Collocation density sensitivity}
\label{sec:collocation}

This subsection examines whether the NTK benchmark errors at
$N_\mathrm{int} = 16{,}000$ are sensitive to the interior collocation budget,
and whether the reported RMSE values are collocation-converged.

\begin{figure}[H]
    \centering
    \includegraphics[width=0.65\textwidth]{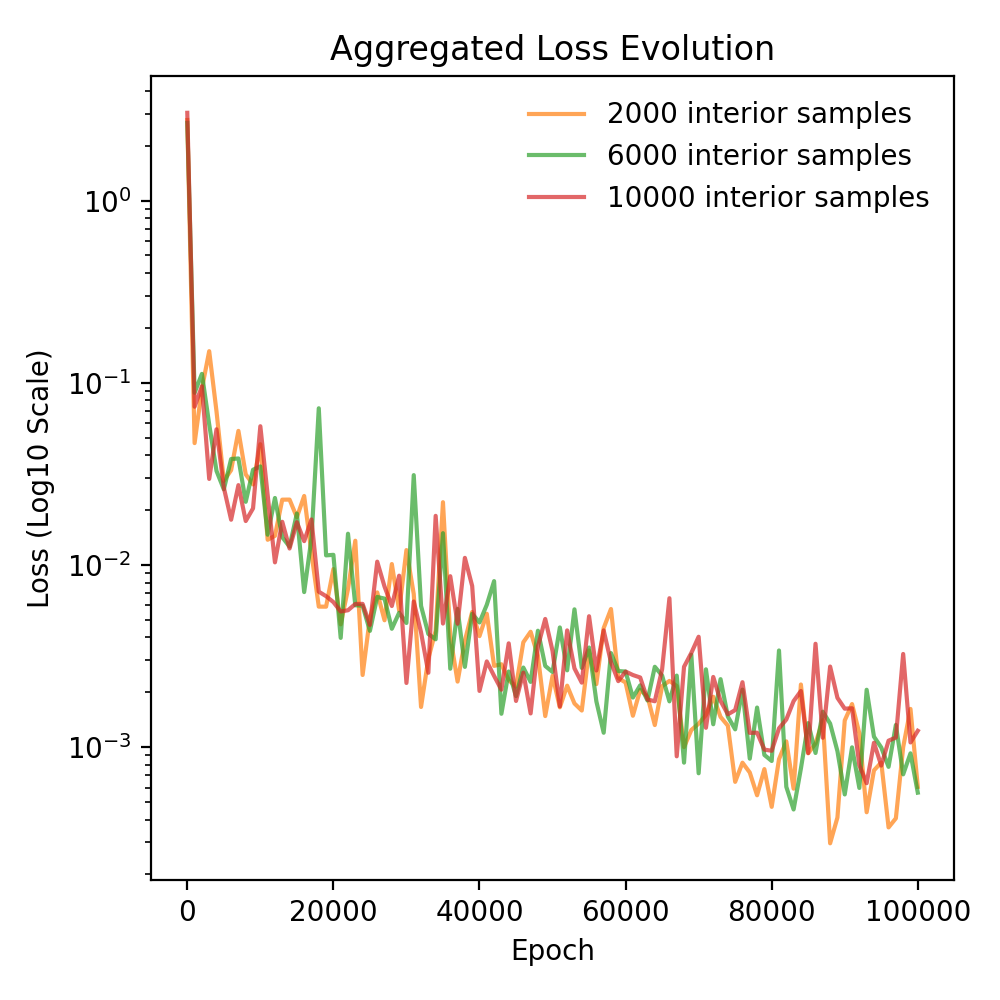}
    \caption{\textbf{NTK training loss for three collocation densities.}
    Aggregated training loss (log$_{10}$ scale) over 100\,000 epochs for the
    NTK-weighted configuration at interior collocation counts
    $N_\mathrm{int} \in \{2{,}000,\,6{,}000,\,10{,}000\}$ (10:1
    interior-to-boundary ratio throughout).
    All three runs converge to a comparable final training loss of
    approximately $10^{-3}$; point-wise RMSE against the FVM reference
    decreases monotonically with increasing $N_\mathrm{int}$.\label{fig:ntk_collocation}}
\end{figure}

Three additional single-seed NTK runs were performed at
$N_\mathrm{int} \in \{2{,}000,\,6{,}000,\,10{,}000\}$, maintaining the 10:1
interior-to-boundary ratio used throughout the benchmark.
Figure~\ref{fig:ntk_collocation} shows the aggregated training loss for all
three runs.

All three runs converge to a final training loss of approximately $10^{-3}$
within the 100\,000-epoch budget, and the loss curves are qualitatively
similar across the three $N_\mathrm{int}$ values.
Inspection of the validator surfaces suggests that point-wise RMSE against the
FVM reference decreases monotonically as $N_\mathrm{int}$ increases from
2\,000 to 10\,000, consistent with the expectation that denser collocation
better resolves the stiff electric double layer near the electrode interfaces.


\section{Conclusions}
\label{sec:conclusions}

This benchmark establishes that adaptive loss weighting is the dominant factor
governing PINN accuracy on the stiff PNP system, outweighing architecture choice,
input-space encoding, and variable reformulation.
NTK weighting achieves dimensionless RMSE as low as $(6.6\pm0.4)\times10^{-4}$
(anion), $(6.2\pm0.3)\times10^{-4}$ (cation), and $(1.1\pm0.1)\times10^{-3}$
(electric potential), establishing the strongest data-free accuracy reported for
a 1D PNP system under battery-relevant parametrisation~\cite{HUANG2025231}.
The cheaper BRDR scheme matches NTK within 10\% of RMSE for the concentration
fields (24\% for $\varphi$) by computing only scalar
residual statistics rather than full NTK matrix traces, reducing mean wall time by
$3.2\pm0.4$\,h per run on an NVIDIA H100; NTK is the right choice when accuracy
is the priority, BRDR when compute is constrained.
Spectral-bias-aware architectures (Fourier features, FBPINN, PIKAN) produce more
spatially uniform error distributions but do not achieve the lowest total RMSE
within the fixed 100\,000-epoch budget, indicating a convergence-speed trade-off
that a precision-target or curriculum training protocol could resolve.
Loss basin sharpness correlates monotonically with the RMSE ranking across all
eleven configurations, providing an efficient geometry-based diagnostic that does
not require FVM comparison.
These findings generalise beyond the specific electrochemical application: the
stiffness structure (small singular-perturbation parameter, inter-equation loss
imbalance) is shared by semiconductor drift-diffusion, reactive porous-media
transport, and coupled thermo-mechanical problems, and the same adaptive-weighting
remedies are expected to transfer.


\section*{Acknowledgements}

C.G.T.F. acknowledges the support of the Natural Sciences and Engineering Research Council of Canada (NSERC) [RGPIN-2024-03989]. This research was enabled in part by support provided by SHARCNET and the Digital Research Alliance of Canada.

%

\section*{Declaration of Competing Interests}

The authors declare that they have no known competing financial interests or personal
relationships that could have appeared to influence the work reported in this paper.

\section*{Data availability}

The full PhysicsNeMo Sym PINN implementation, including all configuration files
and post-processing scripts, is publicly available at \url{https://github.com/Feugmo-Group/physicsnemo-sym-pnp}.

\bibliographystyle{plainnat}
\bibliography{refs}

\end{document}